\documentclass[journal=jpccck,manuscript=article]{achemso}

\setkeys{acs}{doi=true}

\SectionNumbersOn
\usepackage{amsmath}
\newcommand*{\dif}{\mathop{}\mathrm{d}}

\usepackage{graphicx}
\graphicspath{{figures/}}

\usepackage{siunitx}
\DeclareSIUnit\atomicmassunit{u}
\DeclareSIUnit\angstrom{\text{Å}}
\DeclareSIUnit\hartree{\text{\ensuremath{E}}_{\mathrm{h}}}

\usepackage{braket}
\usepackage[version=4]{mhchem}
\usepackage[capitalise]{cleveref}

\usepackage{acronym}
\newacro{CDFT}{constrained density functional theory}
\newacro{IESH}{independent electron surface hopping}
\newacro{TSH}{trajectory surface hopping}
\newacro{NAH}{Newns--Anderson Hamiltonian}
\newacro{AFSSH}{augmented FSSH}
\newacro{EDC}{energy decoherence correction}
\newacro{BCME}{broadened classical master equation}
\newacro{HQME}{hierarchical quantum master equations}
\newacro{MDEF}{molecular dynamics with electronic friction}
\newacro{PES}{potential energy surface}
\newacro{MQC}{mixed quantum-classical}
\newacro{CME}{classical master equation}
\newacro{MD}{molecular dynamics}
\newacro{DFT}{density functional theory}
\newacro{DOS}{density of states}

\author{James Gardner}
\author{Scott Habershon}
\author{Reinhard J. Maurer}
\email{r.maurer@warwick.ac.uk}
\affiliation[University of Warwick]
{Department of Chemistry, University of Warwick, Gibbet Hill Road, Coventry CV4 7AL, United Kingdom}
\alsoaffiliation[University of Warwick]
{Department of Physics, University of Warwick, Gibbet Hill Road, Coventry CV4 7AL, United Kingdom}

\title{Assessing Mixed Quantum-Classical Molecular Dynamics Methods for Nonadiabatic Dynamics of Molecules on Metal Surfaces}

\begin{document}

\begin{tocentry}
\includegraphics{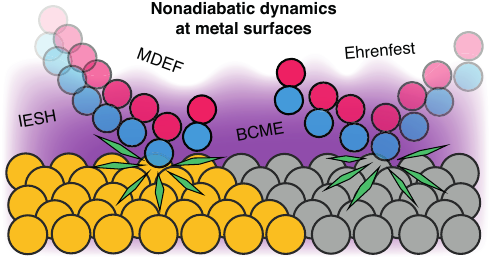}
\end{tocentry}

\begin{abstract}
    Mixed-quantum classical (MQC) methods for simulating the dynamics of molecules at metal surfaces have the potential to accurately and efficiently provide mechanistic insight into reactive processes.
    Here, we introduce simple two-dimensional models for the scattering of diatomic molecules at metal surfaces based on recently published electronic structure data. We apply several MQC methods to investigate their ability to capture how nonadiabatic effects influence molecule-metal energy transfer during the scattering process. Specifically, we compare molecular dynamics with electronic friction, Ehrenfest dynamics, Independent Electron Surface Hopping, and the Broadened Classical Master Equation approach. In the case of Independent Electron Surface Hopping, we implement a simple decoherence correction approach and assess its impact on vibrationally-inelastic scattering. Our results show that simple, low-dimensional models can be used to qualitatively capture experimentally observed vibrational energy transfer and provide insight into the relative performance of different MQC schemes. We observe that all approaches predict similar kinetic energy dependence, but return different vibrational energy distributions. Finally, by varying the molecule-metal coupling, we can assess the coupling regime in which some MQC methods become unsuitable.
\end{abstract}

\section{Introduction}

When atoms and molecules adsorb and react at metal surfaces, they lose kinetic energy by directly exciting electron-hole pair excitations in the metal. Several seminal experimental works have shown the strong impact of nonadiabatic coupling and hot electron effects on experimentally measurable signatures of surface chemistry.  \cite{nienhausElectronHolePairCreation1999,wodtkeElectronicallyNonadiabaticInteractions2004, bunermannElectronholePairExcitation2015}  By establishing a deep understanding of nonadiabatic dynamics at metal surfaces, new applications and technologies that utilize nonadiabatic energy transfer can be developed for catalysis and energy conversion, such as light- and hot-carrier-driven chemistry on plasmonic metal nanostructures.
\cite{brongersmaPlasmoninducedHotCarrier2015,zhangPlasmonDrivenCatalysisMolecules2019}
Achieving insight on an atomistic level requires computational simulation methods that can accurately describe nonadiabatic effects during dynamics while scaling efficiently for realistic systems. The simulation of experimentally measurable reaction and scattering probabilities requires statistically significant averages over many tens of thousands of simulation events. Classical \ac{MD} simulations have proven effective at treating systems where the adiabatic approximation is valid, but going beyond \ac{MD} to include nonadiabatic effects is a challenging task. \cite{tullyPerspectiveNonadiabaticDynamics2012,crespo-oteroRecentAdvancesPerspectives2018}
Many approximate methods that treat electrons quantum mechanically and atoms classicaly, so-called \ac{MQC} methods, have been proposed for the description of coupled electron-nuclear dynamics at metal surfaces,
including \ac{MDEF},
\cite{head-gordonMolecularDynamicsElectronic1995,askerkaRoleTensorialElectronic2016,douBornOppenheimerDynamicsElectronic2017,
douUniversalityElectronicFriction2017,douPerspectiveHowUnderstand2018,boxDeterminingEffectHot2021,
martinazzoQuantumDynamicsElectronic2022}
\ac{IESH},
\cite{shenviDynamicalSteeringElectronic2009,shenviNonadiabaticDynamicsMetal2009,shenviNonadiabaticDynamicsMetal2012,
miaoComparisonSurfaceHopping2019, pradhanDetailedBalanceIndependent2022,gardnerEfficientImplementationPerformance2023}
and \ac{CME} surface hopping.
\cite{douFrictionalEffectsMetal2015,douSurfaceHoppingManifold2015,douMolecularElectronicStates2016,
douBroadenedClassicalMaster2016,douElectronicFrictionMetal2017}

An ongoing challenge for simulating nonadiabatic dynamics at surfaces lies in the reliability of different simulation techniques.
\cite{douNonadiabaticMolecularDynamics2020}
Often, it is difficult to know if the simulations are correctly describing reality as accurate reference results are rare. Progress has been made in this area from two directions, namely, 
verifying methods against quantum dynamics for simple analytical model Hamiltonians,
\cite{douBornOppenheimerDynamicsElectronic2017,loaizaBreakdownEhrenfestMethod2018,miaoComparisonSurfaceHopping2019,
douNonadiabaticMolecularDynamics2020}
and comparing the outcomes of high-dimensional simulations, often based on first-principles electronic structure theory, to experimental observations.
\cite{cooperMultiquantumVibrationalExcitation2012,grotemeyerElectronicEnergyDissipation2014,
bunermannElectronholePairExcitation2015,krugerNOVibrationalEnergy2015,
fuchselReactiveNonreactiveScattering2019,
yinStrongVibrationalRelaxation2019,boxDeterminingEffectHot2021} Both approaches have limitations. The former approach may unduly simplify the electronic structure and the influence of many coupled, anharmonic degrees of freedom. The latter makes it difficult to disentangle errors that arise from the electronic structure description and errors that are intrinsic to the approximations of the applied \ac{MQC} method. For example, \citet{shenviDynamicalSteeringElectronic2009} have applied the \ac{IESH} method to study nonadiabatic vibrational energy loss during nitrous oxide (NO) scattering on Au(111) and they found that the method was able to describe dynamical steering effects connected to vibrational energy loss. Later, it was shown that the employed potential energy landscape based on \ac{DFT} misrepresented energy barriers, which led to an incorrect description of the translational energy dependence of vibrational inelasticity during scattering. \cite{yinStrongVibrationalRelaxation2019}  While previous works questioned the ability of \ac{MDEF} to describe vibrational energy loss for this system, a new and improved potential energy landscape enabled an accurate description with \ac{MDEF}, at least for the case of low vibrational incidence energy. \cite{boxDeterminingEffectHot2021} Hyperthermal scattering of NO from Au(111) and Ag(111) remains one of the most studied experimental reference systems to understand nonadiabatic effects in surface chemistry. \cite{wagnerVibrationalRelaxationHighly2017,wagnerElectronTransferMediates2019} As a quantum reference, the \ac{HQME} promises a numerically exact treatment of coupled electron-vibrational systems, \cite{tanimuraTimeEvolutionQuantum1989,tanimuraStochasticLiouvilleLangevin2006,
schinabeckHierarchicalQuantumMaster2016,erpenbeckCurrentinducedBondRupture2018,
erpenbeckHierarchicalQuantumMaster2019,erpenbeckCurrentinducedDissociationMolecular2020,
kasparNonadiabaticVibronicEffects2022}
however, the method is currently limited to few degrees of freedom which precludes an extension to large atomistic systems. Without a scalable reference method, it is difficult to bridge the gap between simple models and complex systems, casting continued uncertainty on  the validity of approximate \ac{MQC} methods.

Another limitation when developing \ac{MQC} methods is the lack of model systems that can be related to realistic counterparts. The ubiquitous models introduced by \citet{tullyMolecularDynamicsElectronic1990} have been used countless times in recent decades
to benchmark and compare methods for nonadiabatic dynamics
\cite{sunSemiclassicalInitialValue1997,wuNonadiabaticSurfaceHopping2005,
poirierReconcilingSemiclassicalBohmian2007,dunkelIterativeLinearizedApproach2008,
zimmermannCommunicationsEvaluationNondiabaticity2010,gorshkovSemiclassicalMonteCarloApproach2013,
fengNonadiabaticMolecularDynamics2014,cottonSymmetricalQuasiClassicalSpinMapping2015,
agostiniQuantumClassicalNonadiabaticDynamics2016,gosselCoupledTrajectoryMixedQuantumClassical2018,
baskovImprovedEhrenfestApproach2019}
and have been recently shown to closely relate to real molecular systems.
\cite{ibeleMolecularPerspectiveTully2020}
However, similar models for dynamics at metal surfaces are less widespread.\cite{ouyangSurfaceHoppingManifold2015}
A unified collection of models that are capable of relating to experimentally measurable phenomena would be beneficial for further development of \ac{MQC} methods.

In this work we apply \ac{MDEF}, \ac{IESH}, Ehrenfest dynamics, \ac{BCME} and adiabatic \ac{MD} to two-dimensional model Hamiltonians that describe the 
scattering of diatomic molecules on metal surfaces.
The two models introduced are designed to have a simple analytic form for easy implementation and usage, whilst closely matching recently published ground- and excited-state ab initio \acp{PES} to ensure that the models are physically relevant. \cite{mengPragmaticProtocolDetermining2022} Using these models, we explore the effect of decoherence on molecular scattering as modeled by \ac{IESH} and find
that decoherence can have subtle effects on vibrational energy transfer during molecular scattering. Furthermore, we compare the agreement between the full set of \ac{MQC} methods and determine that all methods capture similar trends in kinetic energy dependence for models that feature realistic model parameters, but deviate in the widths of the vibrational distributions. By exploring models with higher and lower molecule-metal coupling than what is observed experimentally, we identify the limitations of the respective \ac{MQC} methods.

Much previous work has focused on the electron transfer problem in a harmonic double-well, within the wide-band limit, where Marcus theory can be used as a benchmark.
\cite{douSurfaceHoppingManifold2015,miaoComparisonSurfaceHopping2019,pradhanDetailedBalanceIndependent2022}
However, rarely has the case been explored where the molecule-metal coupling depends on the molecular coordinates, where the wide-band limit is less well-defined.\cite{douElectronicFrictionMetal2017}
A key novelty of our new models is that they are inspired by ab initio data in order to capture vibrational de-excitation during nonadiabatic scattering. This allows us to study in greater detail the case where the coupling strength depends on the molecule-metal distance.

The outline of the paper is as follows.
In \cref{sec:theory} we introduce the \ac{NAH} and \ac{MQC} methods used for the simulations.
\Cref{sec:models} presents the parametrization of two models based on the well-studied NO on Au(111) and NO on Ag(111) systems.
\Cref{sec:decoherence} explores the effect of decoherence in \ac{IESH} for scattering problems
and \cref{sec:standard-model-results} compares the performance of the \ac{MQC} methods.
In \cref{sec:extreme-coupling}, the coupling strength is modified to investigate how the performance of each method changes.
\cref{sec:conclusion} closes the paper with our conclusions.

\section{Theory}\label{sec:theory}

\subsection{Newns-Anderson Hamiltonian}\label{sec:model-hamiltonian}

The standard model for nonadiabatic dynamics at metal surfaces is the \ac{NAH},
written as
\begin{equation}
    \hat{H}_\text{NA}(\hat{\mathbf{x}},\hat{\mathbf{p}}) = \sum_\nu \frac{\hat{p}_\nu^2}{2m_\nu} + U_0(\hat{\mathbf{x}}) + \hat{H}_\text{el}(\hat{\mathbf{x}}),
    \label{eq:NA-hamiltonian}
\end{equation}
where $\hat{\mathbf{x}}$ is the vector of nuclear coordinate operators and $\hat{\mathbf{p}}$ their conjugate momenta with particle masses $\mathbf{m}$.
The index $\nu$ is used to label each nuclear degree of freedom in the system.
The electronic state-independent potential energy function $U_0$ and the electronic Hamiltonian $\hat{H}_\text{el}$ determine the potential energy of the system.
The electronic Hamiltonian for a discretized metallic continuum of states is
\begin{equation}
    \hat{H}_\text{el}(\hat{\mathbf{x}}) = h(\hat{\mathbf{x}})\hat{d}^\dag \hat{d}
    + \sum_{k=1}^{M} \epsilon_k \hat{c}_k^\dag \hat{c}_k
    + \sum_{k=1}^{M} V_k(\hat{\mathbf{x}}) (\hat{d}^\dag \hat{c}_k + \hat{c}_k^\dag \hat{d})
    \label{eq:discrete-electronic-hamiltonian}
\end{equation}
where $\hat{d}^\dag (\hat{d})$ are the creation (annihilation) operators for an electron in the molecular state and
$\hat{c}_k^\dag (\hat{c}_k)$ are the creation (annihilation) operators for an electron in metal state $k$.
When the molecular state is occupied, $h(\hat{\mathbf{x}}) = U_1(\hat{\mathbf{x}}) - U_0(\hat{\mathbf{x}})$ is added as a further contribution to the system potential energy.
To obtain the coupling terms $V_k$ that allow population transfer between the metal and molecule it is necessary to discretize the hybridization function
\begin{equation}
    \Gamma(\hat{\mathbf{x}},\epsilon) = 2\pi \sum_k |V(\hat{\mathbf{x}})|^2 \delta(\epsilon - \epsilon_k).
\end{equation}
In this work, the problem is simplified using the wide-band approximation such that the hybridization function becomes independent of energy $\Gamma(\hat{\mathbf{x}},\epsilon) = \Gamma(\hat{\mathbf{x}})$.
The coupling terms then become $V_k = w_k\sqrt{\Gamma / 2\pi}$,
where the weights $w_k$ can be obtained using different discretization methods.\cite{gardnerEfficientImplementationPerformance2023}

\subsection{Mixed quantum-classical dynamics methods}\label{sec:dynamics-methods}

\ac{MQC} dynamics methods allow for the simulation of coupled nuclear-electronic dynamics at metal surfaces. The treatment of the nuclei as classical particles ensures their scalability and their computational efficiency, improving their ability to treat complex systems that are not tractable using quantum dynamics methods.
However, using classical nuclei precludes the treatment of nuclear quantum effects. In this manuscript, we only consider classical nuclear motion. The following sections briefly introduce the methods that are used for the simulations in \cref{sec:results}.

\subsubsection{Molecular dynamics with electronic friction}

One of the most popular methods for simulating dynamics at surfaces is \ac{MDEF} which captures electron-nuclear coupling via a system-bath description using a Langevin equation.
\cite{head-gordonMolecularDynamicsElectronic1995,askerkaRoleTensorialElectronic2016,
maurerInitioTensorialElectronic2016,douPerspectiveHowUnderstand2018,martinazzoQuantumDynamicsElectronic2022}
The key ingredient of \ac{MDEF} is the friction tensor
that governs the transfer of energy between the nuclei and electrons.
Although obtaining the friction tensor can be challenging using \textit{ab initio} calculations, \cite{askerkaRoleTensorialElectronic2016, maurerInitioTensorialElectronic2016, boxInitioCalculationElectronphonon2021}
for the \ac{NAH} in the wideband limit the exact friction tensor is given by
\cite{brandbygeElectronicallyDrivenAdsorbate1995,  jinPracticalAnsatzEvaluating2019}
\begin{equation}
    \Lambda_{\nu\mu} = -\pi\hbar\int \dif\epsilon
    \left(
        \partial_\nu h + (\epsilon - h)\frac{\partial_\nu \Gamma}{\Gamma}
    \right)
    \left(
        \partial_\mu h + (\epsilon - h)\frac{\partial_\mu \Gamma}{\Gamma}
    \right)
    A^2(\epsilon)
    \frac{\partial f}{\partial \epsilon}
    \label{eq:exact-friction}
\end{equation}
where
\begin{equation}
    A(\epsilon) = \frac{1}{\pi}\frac{\Gamma/2}{(\epsilon-h)^2 + (\Gamma/2)^2}
\end{equation}
with $\partial_\nu = \partial / \partial x_\nu$ and $\partial f / \partial \epsilon$ the gradient of the Fermi function.

The \ac{MDEF} equations of motion for the \ac{NAH} can be written as
\begin{equation}
\dot{p_\nu} =
- \partial_\nu U_0
- \sum_k^{M+1} \left(
f(\lambda_k) + \frac{\partial f(\lambda_k)}{\partial \lambda_k} \lambda_k 
\right)\partial_\nu \lambda_k
- \sum_\mu \Lambda_{\nu\mu} \frac{p_\mu}{m_\mu}
+ \sum_\mu \sqrt{\frac{2\Lambda_{\nu\mu} m_\mu}{\beta}} \eta_\mu(t),
\label{eq:mdef-eom}
\end{equation}
The first two terms arise due to the adiabatic force where $\{ \lambda_k \}$ are the $M+1$  eigenvalues of the electronic Hamiltonian ($M$ metallic states and 1 molecule state)
and $f(\lambda_k)$ is the Fermi function that ensures a thermal contribution from each state.
The third term is the retarding force that transfers energy from the nuclei to the electrons. 
The fourth and final term is the random force component that ensures the equations of motion correctly recover thermal equilibrium,
where $\eta_\mu(t)$ is a Gaussian-distributed random number with zero mean and unit variance.

\subsubsection{Independent electron surface hopping}

\ac{IESH}\cite{shenviNonadiabaticDynamicsMetal2009,shenviNonadiabaticDynamicsMetal2012,miaoComparisonSurfaceHopping2019,gardnerEfficientImplementationPerformance2023}
models coupled nuclear-electronic dynamics near metal surfaces by coupling a finite set of discretized metallic states with the molecular state. The electrons in the system are propagated independently in time and the coupling between electrons and nuclear degrees of freedom are described via stochastic hops that represent electronic transitions. Previously, \ac{IESH} has been used to investigate the scattering of NO on Au(111),
\cite{shenviDynamicalSteeringElectronic2009,krugerNOVibrationalEnergy2015}
calculate electron transfer rates,\cite{miaoComparisonSurfaceHopping2019}
and describe desorption and scattering in a one-dimensional model.\cite{gardnerEfficientImplementationPerformance2023}
The nuclear dynamics in \ac{IESH} are governed by the Hamiltonian:
\begin{equation}
    H_\text{IESH}(\mathbf{x},\mathbf{p},t) = 
    \sum_{\nu}\frac{p_\nu^2}{2m_\nu}
    + U_0(\mathbf{x})
    + \sum_{k \in \mathbf{s}(t)} \lambda_k(\mathbf{x}),
\end{equation}
where $\mathbf{s}(t)$ is the vector that contains the indices of states occupied by electrons,
such that the summation includes only occupied states.
From this Hamiltonian it is clear that the nuclei evolve on a potential determined by the electronic occupations at each point in time.
The electronic occupations change during the dynamics by allowing a single electron to hop  each time step
with probabilities obtained from Tully's fewest-switches criteria.\cite{tullyMolecularDynamicsElectronic1990}
In order to calculate the hopping probabilities it is necessary to propagate the electronic wavefunctions for each electron alongside the nuclear dynamics
by solving the time-dependent electronic Schr\"odinger equation,
\begin{equation}
    i\hbar \dot{c}_{k} = \lambda_k(\mathbf{x}) c_{k} - i\hbar \sum_j \sum_\nu \frac{p_\nu}{m_\nu} d_{\nu jk}(\mathbf{x}) c_{j},
    \label{eq:electronic_schrodinger_equation}
\end{equation}
where $\set{c_k}$ are the complex expansion coefficients for each electron and $d_{\nu jk}$ is the nonadiabatic coupling along coordinate $\nu$ between adiabatic states $j$ and $k$.
The electronic coefficients are initialized such that they are consistent with the discrete occupations.

\subsubsection{Ehrenfest dynamics}

The Ehrenfest dynamics method allows the nuclei to evolve on the \ac{PES} obtained from the expectation value of the electronic Hamiltonian. 
\cite{ehrenfestBemerkungUeberAngenaeherte1927,mclachlanVariationalSolutionTimedependent1964,tullyPerspectiveNonadiabaticDynamics2012,tullyEhrenfestDynamicsQuantum2023}
The Hamiltonian that describes Ehrenfest dynamics for the nuclei based on an \ac{NAH},  $\hat{H}_{el}$, is
\begin{equation}
    H_\text{Ehr}(\mathbf{x},\mathbf{p},t) = 
    \sum_{\nu}\frac{p_\nu^2}{2m_\nu}
    + U_0(\mathbf{x})
    + \Braket{\psi(t)|\hat{H}_{el}(\mathbf{x})|\psi(t)},
\end{equation}
where $\psi(t)$ is the electronic wavefunction at time $t$.
As with \ac{IESH}, the electronic wavefunction is coherently propagated alongside the nuclear dynamics using \cref{eq:electronic_schrodinger_equation}.
However, unlike both \ac{MDEF} and \ac{IESH},  the Ehrenfest method is entirely deterministic such that each trajectory is uniquely determined by its initial conditions.

\subsubsection{Broadened classical master equation}\label{sec:bcme}

Another alternative is to model the presence of the electronic bath implicitly by representing
the dynamics with a classical master equation that describes the time evolution of the nuclear probability density of the system.
\cite{douFrictionalEffectsMetal2015,douSurfaceHoppingManifold2015}
The \ac{CME} method involves classical dynamics on a single diabatic state,
with transitions between states that ensure the correct thermal equilibrium is reached when $\Gamma$ is small.
The original limitation to the regime of small $\Gamma$ was due to the neglect of broadening effects induced by the molecule-metal coupling.
To go beyond the regime of small $\Gamma$, broadening effects were incorporated by extrapolating the \ac{CME} forces to the adiabatic regime.
The \ac{BCME} recovers the original \ac{CME} when $\Gamma$ is small,
but yields adiabatic dynamics on a broadened potential of mean force when $\Gamma$ is large.
\cite{douBroadenedClassicalMaster2016,douElectronicFrictionMetal2017}
In addition to the modified force, the original proposal for \ac{BCME} included momentum jumps in the algorithm.\cite{douBroadenedClassicalMaster2016}
However, later an alternative form was introduced with slightly modified forces that no longer required any momentum jumps.\cite{douElectronicFrictionMetal2017}

The updated form of the broadened master equation is\cite{douNonadiabaticMolecularDynamics2020}
\begin{align}
    \frac{\partial \rho_0}{\partial t} =&
    - \sum_\nu \frac{p_i}{m_i}\partial_\nu \rho_0
    + \sum_\nu \partial_\nu \tilde{U}_0 \partial_\nu \rho_0
    - \frac{\Gamma}{\hbar} f(h) \rho_0
    \nonumber
    \\
    &+ \frac{\Gamma}{\hbar} (1 - f(h)) \rho_1
    \\
    \frac{\partial \rho_1}{\partial t} =&
    - \sum_\nu \frac{p_i}{m_i}\partial_\nu \rho_1
    + \sum_\nu \partial_\nu \tilde{U}_1 \partial_\nu \rho_1
    + \frac{\Gamma}{\hbar} f(h) \rho_0
    \nonumber
    \\
    &- \frac{\Gamma}{\hbar} (1 - f(h)) \rho_1,
\end{align}
where
\begin{equation}
    \partial_\nu \tilde{U}_k =
    \partial_\nu U_k
    + \left(n_1(h) - f(h)\right) \partial_\nu h
    + n_2(h) \frac{\partial_\nu \Gamma}{\Gamma}
    \label{eq:bcme_force}
\end{equation}
and
\begin{align}
    n_1(h) &= \int_{\infty}^{\infty}\dif\epsilon\, A(\epsilon, h) f(\epsilon)
    \\
    n_2(h) &= \int_{-W/2}^{W/2} \dif\epsilon\, (\epsilon - h) A(\epsilon, h) f(\epsilon).\label{eq:n2}
\end{align}
The two broadening functions $n_1$ and $n_2$ involve a convolution of the Fermi function $f(\epsilon)$
with a Lorentzian function:
\begin{equation}
    A(\epsilon, h) = \frac{1}{\pi}\frac{\Gamma/2}{(\epsilon-h)^2 + (\Gamma/2)^2}.
\end{equation}

The final term in \cref{eq:bcme_force} involving $\partial \Gamma/\partial x_i$ was proposed as an additional
contribution to the force that includes non-Condon effects in the BCME dynamics.\cite{douElectronicFrictionMetal2017}
However, the integral in \cref{eq:n2} diverges logarithmically in the wide-band limit where $W \to \infty$.
\cite{douMolecularElectronicStates2016,douElectronicFrictionMetal2017}
Therefore, whenever $\partial \Gamma/\partial x_i$ is non-zero, the potential of mean force will depend on the width of the band.

\section{Results and Discussion}\label{sec:results}

\subsection{Models}\label{sec:models}

The simulations in this paper are focused on two analytic models
that describe the interaction of an NO molecule with two different metal surfaces:
Au and Ag. We only consider two degrees of freedom: The centre of mass distance between molecule and surface $z$ and the intramolecular distance $r$. In our models, we assume that the molecular axis is always aligned perpendicular to the surface and that the N atom always faces down. 
The form of the two diabatic \ac{PES}s are chosen to be
\begin{equation}
U_0(r,z) = V_M\left[r-r_0;\, D_0, a_0\right]
+ \exp\left[-b_0(z-z_0)\right] + c_0\label{eq:u0}
\end{equation}
\begin{equation}
U_1(r,z) =  V_M\left[r-r_1;\, D_1, a_1\right]
+ V_M\left[z-z_1;\, D_2, a_2  \right] + c_1\label{eq:u1}
\end{equation}
where $V_M$ is the Morse potential defined as:
\begin{equation}
V_M(x;\, D, a) = D\left[\exp(-2ax)- 2\exp(-ax) \right].
\end{equation}
The coupling is chosen to be dependent on only the molecule-surface distance, given by
\begin{equation}
V_k(z) = \bar{V}_k
\left[
1 - \tanh\left(z / \tilde{a}\right)
\right].
\label{eq:coupling}
\end{equation}
The decision has been made to restrict  the models to a simple analytic form
to ensure that the models can be easily understood and implemented.
With the functional form of the models established,
it is necessary to choose values for each of the parameters in
\cref{eq:u0,eq:u1,eq:coupling}.
The neutral $U_0$ bond stretching Morse parameters are taken from \citet{laportaVibrationallyResolvedNO2020}.
To ensure that the models best represent the molecule-metal interaction, the remaining parameters for 
$U_0$ and $U_1$ have been fitted to the \ac{DFT} data presented by \citet{mengPragmaticProtocolDetermining2022}. \citet{mengPragmaticProtocolDetermining2022} employed \ac{CDFT} to model the scenario where the molecule does not exchange charge with the surface ($U_0$) and where the molecule accepts a full electron from the surface ($U_1$). They presented the adiabatic and diabatic \ac{PES}s for several one-dimensional curves (Figures 7 and 9 in the paper) along $z$ and $r$ for each of the two metal surfaces, Au(111) and Ag(111), with the molecule laterally placed in an hcp site with the N atom facing down. 
Using the reference data,
\cref{eq:u0,eq:u1} have been fitted using the gradient-free Nelder--Mead method\cite{nelderSimplexMethodFunction1965} as implemented in the \textit{Optim.jl} package.
\cite{mogensenOptimMathematicalOptimization2018,mogensenJuliaNLSolversOptimJl2022}

\begin{figure}
\includegraphics{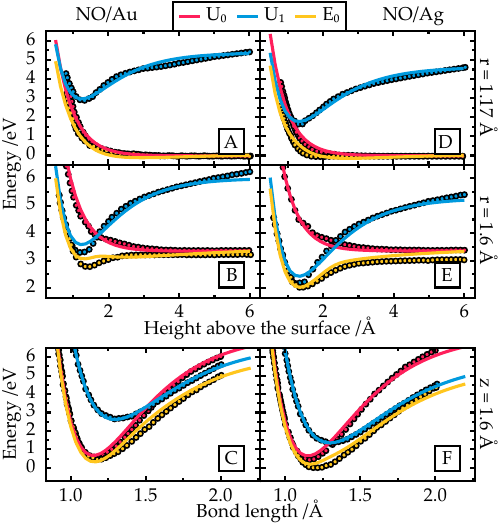}
\caption{One-dimensional slices of the two diabatic \ac{PES}s $U_0, U_1$ and the adiabatic ground state energy $E_0$ for the two models (NO/Au in panels A-C and NO/Ag in panels D-F).
The analytic models are shown with solid lines and the dots show the \ac{DFT} data from Ref.~\citenum{mengPragmaticProtocolDetermining2022} used to create the models.
In panels A, B, D, and E, the molecule has the fixed bond lengths labeled on the right of the figure.
In panels C and F, the molecule has a fixed height above the surface of \SI{1.6}{\angstrom}.
}
\label{fig:model-fitting}
\end{figure}

The resulting $U_0$ and $U_1$ functions are shown in \cref{fig:model-fitting}.
Generally, the choice of functional form appears suitable for capturing the shape of each of the diabatic curves,
although not providing a perfect fit in some areas.
Some significant qualitative differences exist for $U_1$ in panels B and E,
where the depths of the minima are slightly underestimated.
However, our goal is to obtain only a simple, physically motivated model, where a quantitative match of the \ac{DFT} energetics across the \ac{PES} is not required.
An important feature captured by the models is that NO on Ag(111) has a reduced energy gap between the neutral and anionic diabatic states, $U_0$ and $U_1$, compared to Au(111). This is caused by the fact that Ag(111) has a lower work function than Au(111), facilitating the energy transfer from the metal to the molecule. 
This leads to a crossing of the diabats at a reduced bond length for NO on Ag and an enhancement of nonadiabatic electron transfer.
\cite{mengPragmaticProtocolDetermining2022}

The coupling function in \cref{eq:coupling} describes a monotonic decay as the molecule moves away from the surface.
Since there are only two parameters,
their values are simply chosen in order to best recover the adiabatic \ac{DFT} ground state \ac{PES},
also shown in \cref{fig:model-fitting} as a yellow curve.
To keep the models as simple as possible,
the coupling parameters are chosen to be the same for both NO/Ag and NO/Au models.
As with the two diabats, the qualitative agreement between the analytic model and the reference data is good.
The full set of parameters for \crefrange{eq:u0}{eq:coupling} is given in \cref{tab:parameters}.

\begin{figure}
\includegraphics{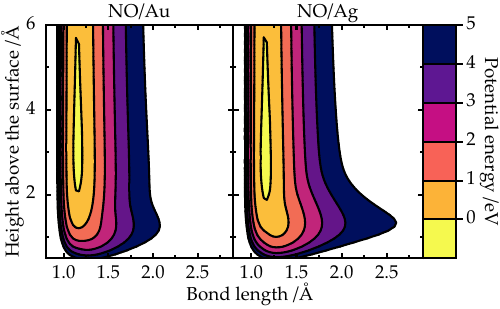}
\caption{NO/Au and NO/Ag adiabatic ground state potential energy surfaces.}
\label{fig:model-pes}
\end{figure}

The resulting ground state \ac{PES}s are shown in \cref{fig:model-pes}.
In the entrance channel, both appear similar,
but as the molecule approaches the surface, there is a softening of the bond stretching potential that is stronger for Ag than for Au.
The surfaces remain comparable at short bond lengths near the surface,
where the neutral state is lower in energy,
but for NO/Ag the softening of the bond stretching potential is more pronounced.
This is consistent with the crossing of the diabatic surfaces at a shorter bond length $r$.

\begin{table}
    \caption{Parameters for the two NO models.
    The values for the NO Morse potential and coupling function are shared by both models.
    \label{tab:parameters}}
    \begin{tabular*}{0.475\columnwidth}{@{\extracolsep{\fill}}ll}
    \hline
    NO Morse&\\
    \hline
    $r_0$ &\SI{1.1510}{\angstrom}\\
    $a_0$ &\SI{2.7968}{\per\angstrom}\\
    $D_0$ &\SI{6.610}{\eV}\\
    \hline
    \end{tabular*}
    \quad
    \begin{tabular*}{0.475\columnwidth}{@{\extracolsep{\fill}}ll}
    \hline
    Coupling &\\
    \hline
    $\Gamma$ & \SI{1.5}{\eV}\\
    $\bar{V}_k$ & $\sqrt{\Gamma / 2\pi}$\\
    $\tilde{a}$ &\SI{10}{\angstrom}\\
    \hline
    \end{tabular*}
    \vspace{1ex}\newline
    \begin{tabular*}{0.475\columnwidth}{@{\extracolsep{\fill}}ll}
    \hline
    NO/Au &\\
    \hline
    $b_0$ &\SI{1.9535}{\per\angstrom}\\
    $z_0$ &\SI{-0.26876}{\angstrom}\\
    $c_0$ &\SI{6.5713}{\eV}\\
    $a_1$ &\SI{2.5194}{\per\angstrom}\\
    $r_1$ &\SI{1.2950}{\angstrom}\\
    $D_1$ &\SI{4.1528}{\eV}\\
    $a_2$ &\SI{1.0015}{\per\angstrom}\\
    $z_1$ &\SI{1.2350}{\angstrom}\\
    $D_2$ &\SI{2.4171}{\eV}\\
    $c_1$ &\SI{8.9587}{\eV}\\
    \hline
    \end{tabular*}
    \quad
    \begin{tabular*}{0.475\columnwidth}{@{\extracolsep{\fill}}ll}
    \hline
    NO/Ag &\\
    \hline
    $b_0$ &\SI{2.0402}{\per\angstrom}\\
    $z_0$ &\SI{-0.21164}{\angstrom}\\
    $c_0$ &\SI{6.5804}{\eV}\\
    $a_1$ &\SI{2.4062}{\per\angstrom}\\
    $r_1$ &\SI{1.2963}{\angstrom}\\
    $D_1$ &\SI{4.5879}{\eV}\\
    $a_2$ &\SI{0.92289}{\per\angstrom}\\
    $z_1$ &\SI{1.3161}{\angstrom}\\
    $D_2$ &\SI{2.8481}{\eV}\\
    $c_1$ &\SI{8.6327}{\eV}\\
    \hline
    \end{tabular*}
\end{table}

\subsection{Computational details}
The results presented in \cref{sec:standard-model-results,sec:decoherence,sec:extreme-coupling}
are obtained from molecular scattering simulations using the models in \cref{sec:models}
and the methods introduced in \cref{sec:dynamics-methods}.
The masses associated which each degree of freedom corresponded to the physical mass of the NO molecule;
the reduced mass was used for the bond stretching motion and the total mass for the translation along the surface-adsorbate distance.
In all cases, 2000 trajectories were used for every incidence kinetic energy and vibrational initial state.
The molecule begins at a height of \SI{5}{\angstrom} in a given vibrational state initialized
using the Einstein-Brillouin-Keller semi-classical quantization method for a diatomic molecule
as described in Ref.~\citenum{larkoskiNumericalImplementationEinsteinBrillouinKeller2006}.
The translational velocity is set corresponding to a given kinetic energy $E_i / \si{\eV} \in [0.2, 1.0]$.
The electronic temperature is set to \SI{300}{\kelvin} and a timestep of \SI{0.25}{\femto\second} is used.
The metallic bath comprises 200 states with a bandwidth of \SI{100}{\eV}.
The bandwidth was chosen to be sufficiently wide to ensure the models exist in the wideband limit.
After selecting the bandwidth, the number of metal states was increased until convergence was obtained. This procedure has previously been described. \cite{gardnerEfficientImplementationPerformance2023}

Initializing the electronic state requires special consideration for each method.
For \ac{IESH}, the initial electronic populations are sampled to be consistent with the Fermi-Dirac distribution at the given electronic temperature.
The Ehrenfest simulations are initialized in a similar way,
where the electronic wavefunction is initially consistent with sampled discrete electronic populations.
Unlike \ac{IESH} and Ehrenfest, \ac{BCME} is propagated in the diabatic representation.
Therefore the \ac{BCME} simulations are initialized such that the molecular level is unpopulated. 
In the case of \ac{MDEF}, the electronic populations are simply governed by the Fermi function.

Simulations are terminated when the molecular center of mass exceeds \SI{5}{\angstrom}
or the duration of the simulation reaches \SI{1}{\pico\second}.
Final vibrational states are obtained using the reverse of the initial quantization procedure.
In the case that the time limit is reached, the trajectory is excluded from any vibrational analysis.
The standard error in each probability value is calculated as $\sqrt{p_i(1-p_i)/N}$
where $p_i$ is each individual probability and $N$ is the total number of trajectories that scatter.
All simulations were carried out using the open-source software package \textit{NQCDynamics.jl} v0.13.3.\cite{gardnerNQCDynamicsJlJulia2022}
The default integration algorithms within \textit{NQCDynamics.jl} were used for all methods.
For \ac{MDEF} this is the BAOAB algorithm,
\cite{leimkuhlerRobustEfficientConfigurational2013}
for \ac{IESH} and Ehrenfest the augmented Verlet algorithm as described previously,
\cite{gardnerEfficientImplementationPerformance2023}
and for \ac{BCME} the adaptive 5th order Adams-Bashforth-Moulton method (VCABM5).
\cite{hairerSolvingOrdinaryDifferential1993,rackauckasDifferentialEquationsJlPerformant2017}
The adaptive method used the same \SI{0.25}{\femto\second} as the initial time step,
with absolute and relative error tolerances set equal to \num{1e-10} in atomic units.

\subsection{Decoherence in IESH}\label{sec:decoherence}
\Ac{TSH} simulations suffer from the issue of overcoherence,
where the coherent propagation of the electronic wavefunction becomes inconsistent
after bifurcation of the nuclear wavepacket.
\cite{tullyMolecularDynamicsElectronic1990,subotnikUnderstandingSurfaceHopping2016,
plasserStrongInfluenceDecoherence2019}
To address the issue, a collection of algorithmic modifications have been proposed,
collectively referred to as decoherence corrections.
\cite{subotnikFewestSwitchesSurfaceHopping2011}
These involve adapting the coherent propagation of the electronic wavefunction to improve
the internal consistency between the nuclear and electronic subsystems.
Note also that coherence is not an issue restricted to surface hopping methods,
affecting other methods including Ehrenfest dynamics.
Recently a branching correction has been proposed that can be used to improve both surface hopping and mean-field methods.
\cite{xuBranchingCorrectedSurface2019,xuBranchingCorrectedMean2020,liUnifiedFrameworkMixed2022}
For \ac{IESH}, the importance of decoherence has previously been assessed by comparing rates
and diabatic populations from decoherence-corrected \ac{IESH} and Marcus theory.\cite{pradhanDetailedBalanceIndependent2022}
By adapting the \ac{AFSSH} decoherence correction
\cite{subotnikNewApproachDecoherence2011,jainEfficientAugmentedSurface2016}
for \ac{IESH}, it was shown that
treatment of decoherence improves the simulation results of \ac{IESH} by more accurately preserving detailed balance.\cite{pradhanDetailedBalanceIndependent2022}

\begin{figure}
    \includegraphics{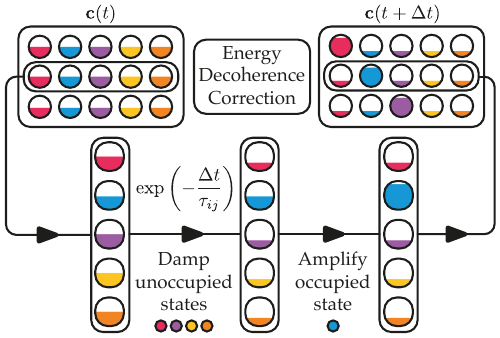}
    \caption{
        Graphical depiction of the \ac{EDC} method for incorporating decoherence into the IESH algorithm using \cref{eq:tau,eq:scale1,eq:scale2}.
        The colored objects represent the wavefunction coefficient for each electron in each basis state,
        in the diagram, there are three electrons with five basis states, each state has a different color.
        The fraction of the object that is colored represents the magnitude of the coefficient. 
    }
    \label{fig:edc}
\end{figure}

In this section, the simple \ac{EDC} method
\cite{granucciCriticalAppraisalFewest2007,granucciIncludingQuantumDecoherence2010}
is adapted for \ac{IESH} and its effect on the vibrational state-to-state scattering probabilities
for the NO models introduced in \cref{sec:models} is explored.
The \ac{EDC} method defines a decoherence time between electronic states $i$ and $j$
\begin{equation}
    \tau_{ij} = \frac{\hbar}{|\lambda_i - \lambda_j|}\left(1 + \frac{C}{E_\text{kin}}\right),
    \label{eq:tau}
\end{equation}
where $E_\text{kin}$ is the kinetic energy and $C$ is a parameter set to \SI{0.1}{\hartree}.\cite{zhuCoherentSwitchingDecay2004}
At every step, $\tau_{ij}$ is used to damp the coefficients of the unoccupied states $c_i$ with
\begin{equation}
    c_i(t + \Delta t) = c_i(t) \exp\left(-\frac{\Delta t}{\tau_{ij}}\right),
    \label{eq:scale1}
\end{equation}
preserving the norm of the wavefunction by increasing the coefficent of the occupied state $c_j$ as
\begin{equation}
    c_j(t + \Delta t) = c_j(t)
    \left[
        \frac{1 - \sum_{i\neq j} |c_i(t+\Delta t)|^2}{|c_j(t)|^2}
    \right]^{1/2}.
    \label{eq:scale2}
\end{equation}
This procedure can be extended for \ac{IESH} by simply repeating the coefficient scaling for each electron in turn,
such that \cref{eq:scale1,eq:scale2} are applied to the individual wavefunctions, using the occupations of each electron.
The full version of \ac{EDC} adapted for \ac{IESH} is depicted in \cref{fig:edc}.
The diagram shows how a single electron wavefunction is selected from $\mathbf{c}(t)$
and how the two operations that make up the \ac{EDC} method are applied in turn to give the decoherence corrected wavefunction.
In \cref{fig:edc}, \cref{eq:tau,eq:scale1} are applied, reducing the magnitude of coefficients for the unoccupied states,
then the occupied state is amplified using \cref{eq:scale2}.
This procedure is repeated for each of the single electron wavefunctions.

\begin{figure}
\includegraphics{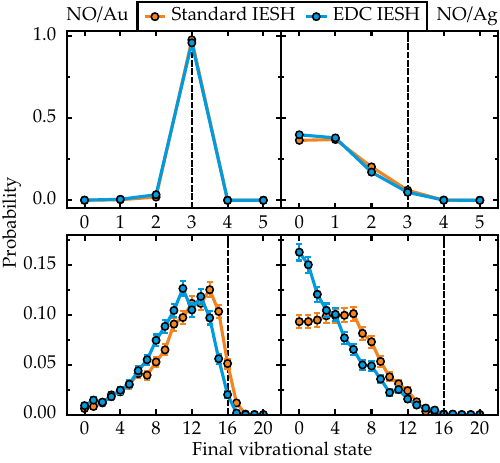}
\caption{
    Final vibrational state probability distributions for both NO models
    (Au left column, Ag right column) using \ac{IESH} with and without a decoherence correction.
    Results are shown for two initial vibrational states $\nu_i \in (3, 16)$ (top and bottom panel, respectively) with an
    incidence energy of \SI{0.2}{\eV}.
    The vibrational initial state is indicated by the vertical dashed lines.
    The error bars show the standard error associated with each point. 
}
\label{fig:decoherence-low-incidence}
\end{figure}

\begin{figure}
\includegraphics{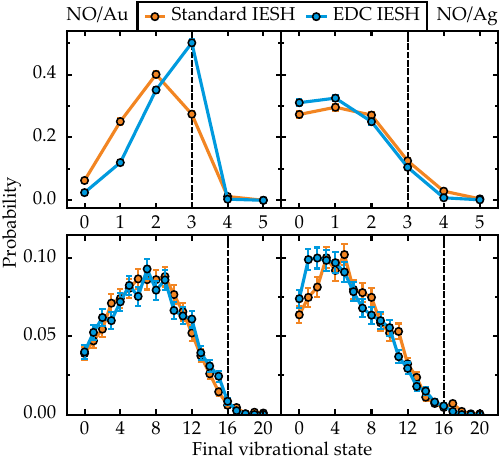}
\caption{
    Final vibrational state probability distributions presented as in \cref{fig:decoherence-low-incidence}
    but here with an incidence energy of \SI{1.0}{\eV}.
}
\label{fig:decoherence-high-incidence}
\end{figure}

The effect of including a decoherence correction within \ac{IESH} is illustrated
by the final state distributions presented in
\cref{fig:decoherence-low-incidence,fig:decoherence-high-incidence}.
In \cref{fig:decoherence-low-incidence},
the distributions obtained with low incidence energy (\SI{0.2}{\eV}) are shown.
When the vibrational energy is low ($\nu_i = 3$) as in the top row,
the decoherence correction has little effect on the final state distribution.
However, with high vibrational energy ($\nu_i = 16$),
including the decoherence correction changes the shape of the final state distributions.
For the Au model, the peak of the distribution is shifted towards lower vibrational states. This also eliminates the small population of scattering events that have led to vibrational excitation from $\nu_i = 16$ to $\nu_f = 17$.
In contrast, for the Ag model, vibrational de-excitation is enhanced such that
the lowest energy states are the most populated.

The corresponding results for a high translational incidence energy of \SI{1.0}{\eV} are shown in \cref{fig:decoherence-high-incidence}.
This time, even with low vibrational energy,
the decoherence correction changes the final state distribution.
The most significant change is observed for Au where the vibrational de-excitation is reduced and the probability of vibrationally elastic scattering is considerably increased.
In the other three cases, the effect is more subtle, only slightly adjusting the individual probabilities. 
In the case of high vibrational energy for Ag (lower right panel),
the effect of \ac{EDC} appears similar to that observed at low incidence,
where the population of intermediate states ($\approx 8$) is reduced and for the lowest energy states is increased.
Overall, the decoherence correction appears to reduce nonadiabatic vibrational inelasticity for NO on Au while increasing it for NO on Ag. However, in general, it is difficult to predict how the effect of the decoherence correction is influenced by the magnitude and partitioning of the initial energy and the specific parameters of the model Hamiltonians.

In comparison with the AFSSH-modified IESH\cite{pradhanDetailedBalanceIndependent2022},
the present \ac{EDC} method has the advantage that it is simple to implement
and has a negligible computational cost.
To incorporate an efficient implementation of \ac{AFSSH} decoherence within \ac{IESH}, additional approximations to the standard method are necessary,
however, for \ac{EDC} it is possible to directly use the standard algorithm without modification.
In the future, it would be interesting to compare the performance of the different decoherence corrections for \ac{IESH} in terms of both accuracy and computational efficiency.

The results in \cref{fig:decoherence-low-incidence,fig:decoherence-high-incidence}
suggest that for these models the effect of decoherence is relatively subtle,
but can lead to quantitative deviations in the result,
particularly when the energy of the projectile is high.
As such, to obtain a fair comparison with the other methods,
all \ac{IESH} results in the subsequent sections will include the \ac{EDC} modification.

\subsection{Comparison of mixed quantum-classical methods}\label{sec:standard-model-results}

In this section, the full collection of methods introduced in \cref{sec:theory} are
applied to the models introduced in \cref{sec:models}.
The goal of these simulations is to identify how effectively each method performs in the prediction of vibrationally inelastic scattering.
However, in lieu of an exact quantum reference, it is difficult to know which method is performing best.
For similar systems where the wide band limit approximation is applied, it has been shown that \ac{BCME} is able to closely reproduce the exact \ac{HQME} result where quantum nuclear effects do not play a role.
\cite{douBroadenedClassicalMaster2017}
With this in mind, although not a perfect reference, we consider \ac{BCME} as a meaningful reference to comparatively assess the performance of the other methods.

When discussing the expected performance of approximate methods for coupled molecule-metal systems,
it is possible to use simple attributes of the model to estimate whether nonadiabatic effects will be significant,
and which methods will be most reliable.
For example, the relevant quantities are often the thermal energy $k_BT$, the molecule-metal coupling strength $\Gamma$,
and for a harmonic system, $\hbar\omega$, which provides a measure for the timescale of nuclear motion in a potential well.
Comparing these quantities allows for the model to be classified and conclusions to be drawn regarding the effectiveness of each method.\cite{douNonadiabaticMolecularDynamics2020}
However, the use of the thermal energy $k_BT$ and nuclear frequency $\hbar\omega$ requires that the system be at thermal equilibrium,
which is not the case during scattering simulations.
Furthermore, when $\Gamma$ depends on the position of the adsorbate, a straightforward comparison is no longer possible.

\begin{figure}
    \includegraphics{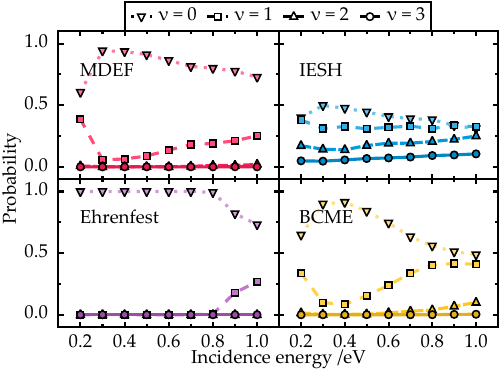}
    \caption{
        Final vibrational state probabilities as a function of incidence energy for the NO/Ag model with $\nu_i = 3$.
        Results are shown for \ac{MDEF}, \ac{IESH}, Ehrenfest and \ac{BCME}.
        The final vibrational state probabilities are shown with markers as indicated by the legend,
        the corresponding lines join the markers to better illustrate the trends.
        The colors are used to identify each of the methods.
        Statistical error bars are not shown as the error is too small to be visible.
    }
\label{fig:vibrational_distribution_NOAg_3}
\end{figure}

\begin{figure}
    \includegraphics{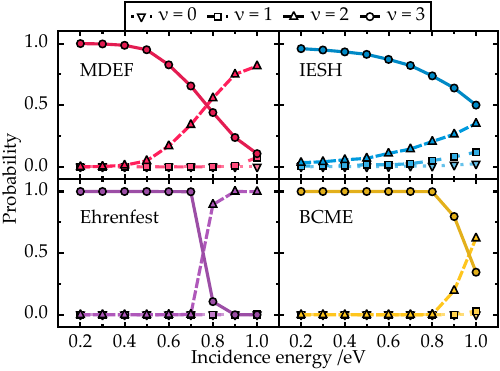}
    \caption{
        Final vibrational state probabilities as a function of incidence energy for the NO/Au model with $\nu_i = 3$.
        Results are presented as in \cref{fig:vibrational_distribution_NOAg_3}.
    }
\label{fig:vibrational_distribution_NOAu_3}
\end{figure}

To understand how each of the methods performs,
simulations have been carried out for high and low initial vibrational states with $\nu_i \in (3, 16)$
as a function of translational incidence energy.
The final state distributions for $\nu_i = 3$ are shown in
\cref{fig:vibrational_distribution_NOAg_3,fig:vibrational_distribution_NOAu_3}.
We do not show the results of the adiabatic simulations as they are entirely vibrationally elastic for both models.
As such, in the case of the low-dimensional models discussed here,
nonadiabatic coupling is solely responsible for all vibrational deexcitation in the following results.
Note that in realistic high-dimensional gas-surface dynamics, vibrational inelasticity can also occur simply due to the anharmonicity of the \ac{PES} and the coupling with the substrate phonons.
In most cases, all trajectories scatter successfully within the \SI{1}{\pico\second} simulation time limit,
however, for some parameter combinations a small fraction remains trapped on the surface.
The proportion of these trapped trajectories is small enough ($< 0.01\%$) to be regarded as negligible.

Considering first the results for the NO/Ag model in \cref{fig:vibrational_distribution_NOAg_3},
it is observed that the dominant final state is $\nu = 0$ across all incidence energies from \qtyrange{0.2}{1.0}{\eV}.
This corresponds to a significant loss in vibrational energy.
The results for \ac{MDEF} and \ac{BCME} appear most similar,
with an initial increase in deexcitation from \SI{0.2}{\eV} to \SI{0.3}{\eV}
followed by a gradual decrease as the incidence energy continues to increase.
Of the four methods, \ac{IESH} is the most distinct, returning a broad distribution
that is largely independent of incidence energy in all four channels.
Most notably, \ac{IESH} is the only method that gives a non-zero probability for the vibrationally elastic channel with a probability of $\approx 0.1$.
On the other hand, Ehrenfest dynamics yield a vibrationally cold final state of $\nu = 0$ up to \SI{0.9}{\eV}. Compared to all other methods, Ehrenfest appears to heavily overestimate nonadiabatic energy loss in this case.

For the NO/Au model (\cref{fig:vibrational_distribution_NOAu_3}), at low incidence energy the scattering is vibrationally elastic,
and as the incidence energy increases the $\nu=2$ channel gains probability,
becoming the dominant channel at high incidence for all methods except \ac{IESH}.
As with the NO/Ag model, the \ac{MDEF} and \ac{BCME} results are most similar but here \ac{MDEF}
shows a smoother transition from $\nu=3$ to $\nu=2$ that occurs at a lower incidence kinetic energy.
Ehrenfest shows a sharp transition similar to \ac{BCME} but the crossover is shifted to lower incidence energy by approximately \SI{0.2}{\eV}.
Again \ac{IESH} is the most unique, showing a gentler transition than \ac{MDEF}.
It is also the only method that shows a significant probability for multi-quantum energy loss with a small probability for $\nu=1$ for incidence energies higher than \SI{0.7}{\eV}.
The NO on Au(111) system has been investigated both experimentally and theoretically in the past for this choice of vibrational state ($\nu_i=3$).
\cite{golibrzuchImportanceAccurateAdiabatic2014}
Although a quantitative agreement is not expected due to the low dimensionality and approximate nature of the current model,
we find that the kinetic energy trends in \cref{fig:vibrational_distribution_NOAu_3} are consistent with the experimental result.
Compared to the experimental data in Fig.~3 of Ref.~\citenum{golibrzuchImportanceAccurateAdiabatic2014}
we see a similar decrease in $\nu_f=3$ and a corresponding increase in $\nu_f=1,2$ probabilities as a function of incidence energy.
The most notable shortcoming of the present results is the overestimation
of the vibrationally elastic channel at lower incidence kinetic energies.
Likely, the low dimensionality of the model that precludes dynamical steering, mode coupling, and phonon-phonon dissipation is responsible for this.

\begin{figure}
\includegraphics{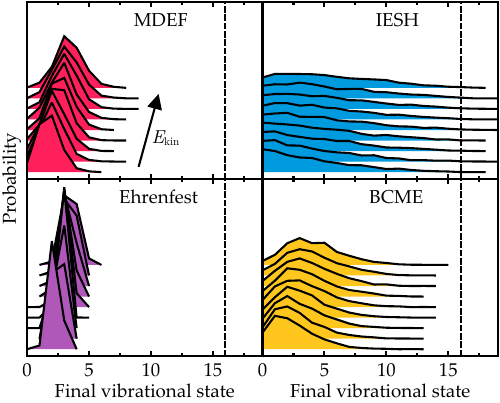}
\caption{
    The final vibrational state distributions as a function of incidence energy as predicted
    by each method for the NO/Ag model.
    The dashed vertical line shows the initial vibrational state ($\nu_i = 16$).
    Distributions of increasing incidence energy $E_\text{kin}$ are stacked on top of each other in the direction of the arrow.
    The incidence values range from \SI{0.2}{\eV} to \SI{1.0}{\eV} in increments of \SI{0.1}{\eV}.
}
\label{fig:vibrational_distribution_NOAg_16}
\end{figure}

\begin{figure}
\includegraphics{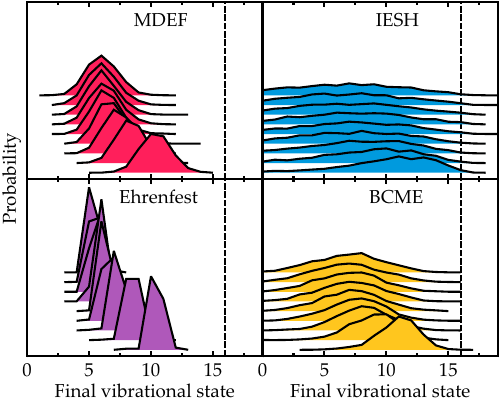}
\caption{
    Data presented as in \cref{fig:vibrational_distribution_NOAg_16} but here for the NO/Au model.
}
\label{fig:vibrational_distribution_NOAu_16}
\end{figure}

In contrast to the low vibrational energy results, molecules prepared 
with high vibrational initial state $\nu_i=16$ yield distributions that are much broader, with final states ranging from 0 up to 20.
To illustrate how the final  vibrational state distribution changes as a function of energy,
the data is presented as a set of probability distributions in \cref{fig:vibrational_distribution_NOAg_16,fig:vibrational_distribution_NOAu_16}.
From these distributions it is possible to see how the centre and shape of the distributions change as a function of translational incidence energy. 

For high vibrational energy $\nu_i=16$ and the NO/Ag model (\cref{fig:vibrational_distribution_NOAg_16}), the results for each method vary significantly,
particularly in terms of the widths of the final state distributions.
As the incidence energy increases all methods follow the same trend,
where the distribution shifts to higher vibrational states, with the shape of the distributions remaining mostly unchanged.
Regarding the width of the distributions,
it appears that the widths increase in the order Ehrenfest $<$ \ac{MDEF} $<$ \ac{BCME} $<$ \ac{IESH}.
For \ac{IESH}, we even find a very small population of trajectories that lead to  vibrational excitation which,
although slightly suppressed by the decoherence correction introduced in \cref{sec:decoherence},
is unexpected for this system when compared to the \ac{BCME} result. The \ac{IESH} method also yields the highest energy loss with the highest point of the distribution positioned at a smaller vibrational final state when compared to the other methods. In contrast, Ehrenfest predicts a very narrow distribution of vibrational states, regardless of the incidence energy.

The results for the NO/Au model with high vibrational energy are shown in \cref{fig:vibrational_distribution_NOAu_16}.
For this model, the incidence energy dependence is opposite to that found for the NO/Ag model, so with increasing kinetic energy, the average final vibrational state goes down and the molecule loses more vibrational energy.
However, as observed for the NO/Ag model, the same trend in distribution widths is observed,
with \ac{IESH} overestimating the \ac{BCME} width and \ac{MDEF} and Ehrenfest underestimating it.
The kinetic energy dependence observed for the two models can be explained with
reference to the diabatic \ac{PES}s in \cref{fig:model-fitting},
specifically panels A,B,D and E,
where the energy is shown as a function of height above the surface.
As pointed out in \cref{sec:models},
nonadiabatic effects are enhanced by the relative alignment of $U_0$ and $U_1$ for Ag(111) compared to Au(111).
However, when the incidence kinetic energy is increased for the NO molecule on Au(111),
the molecule travels closer to the surface,
experiencing enhanced nonadiabatic interaction and corresponding vibrational relaxation.
In contrast, for Ag(111), increasing the kinetic energy only reduces the amount of  time that the molecule
spends in the coupling region.
The key to the different behavior is the alignment of the diabatic surfaces,
where for Ag(111) they cross at a distance further from the surface and at lower energy.

The experimental result of highly vibrationally excited NO scattering on Ag(111) and Au(111) has been investigated previously (\SI{0.14}{\eV} and \SI{0.51}{\eV} incidence energy for $\nu_i=11$ on Ag(111)\cite{krugerVibrationalInelasticityHighly2016}
and \SI{0.5}{\eV} and \SI{1.0}{\eV} for $\nu_i=16$ on Au(111)\cite{krugerNOVibrationalEnergy2015}).
In the case of both Au(111) and Ag(111),
the effect of incidence kinetic energy on the final state distributions appears fairly small.
Although not immediately apparent, this is consistent with the results obtained here.
For Au(111), the most significant change in distribution is observed in the range \qtyrange{0.2}{0.5}{\eV},
above this range the distribution remains relatively unchanged (see \cref{fig:vibrational_distribution_NOAu_16}).
It is exactly in this range where the experimental results are available and the agreement is found.
With Ag(111), the kinetic energy dependence remains constant across the entire range of incidence energies,
but is much less pronounced than in the case of Au(111) (see \cref{fig:vibrational_distribution_NOAg_16}).
Without the experimental results for all incidence energies, it is difficult to conclude whether the model
captures the translational energy dependence,
but with the available data, the agreement appears satisfactory.
When comparing to the experimental results it must be emphasized that the fixed molecular orientation and neglect of surface motion may lead to significant limitations. 
In particular, it has been shown that the initial molecular orientation (N atom facing down or O down)
can influence the observed vibrational energy transfer.
\cite{bartelsObservationOrientationdependentElectron2013,boxDeterminingEffectHot2021}
Any orientation or steering effects\cite{shenviDynamicalSteeringElectronic2009,zhangStereodynamicsAdiabaticNonadiabatic2022}
are clearly neglected by the present two-dimensional models.

Considering the results at high and low vibrational energy for both models,
it appears that \ac{MDEF} predicts results that are in closest agreement with \ac{BCME},
where the average final states are consistently similar.
The most notable shortcoming of \ac{MDEF} lies in underestimating the distribution widths for $\nu_i=16$. This is consistent with what was found   for full-dimensional MDEF simulations of NO scattering on Au(111). \cite{boxDeterminingEffectHot2021}
\ac{IESH} consistently has the opposite problem, overestimating the distribution widths but similarly
capturing the trends in the average final state.
The Ehrenfest method always returns the narrowest distributions, which are clearly inconsistent with experimental findings on the systems.
Both IESH and Ehrenfest are known to suffer from issues related to long-time equilibration\cite{parandekarDetailedBalanceEhrenfest2006,miaoComparisonSurfaceHopping2019}.
However, we do not expect these issues to significantly affect our results as the interaction time during the scattering process is very short.
Any conclusions drawn from these results have the caveat that the molecule-metal coupling strength
is the same in each scenario, namely \SI{1.5}{\eV} at an adsorption height of \SI{0}{\angstrom}, which corresponds to the position of the surface top layer. Only the vibrational and kinetic energies have been varied. It is hard to judge if this is a coupling regime in which all methods can still be considered valid. Therefore, in 
\cref{sec:extreme-coupling}, we explore artificial models with strongly reduced and increased coupling $\Gamma$ to explore the limitations of the respective methods.

\subsection{NO/Au model with extreme coupling values}\label{sec:extreme-coupling}

\begin{figure}
    \includegraphics{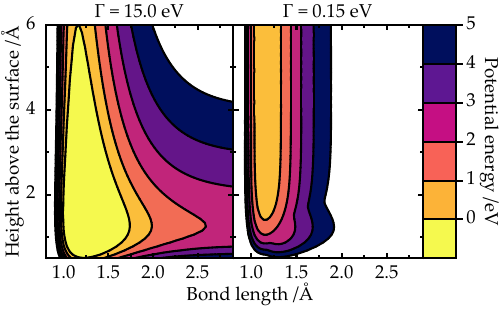}
    \caption{
        Adiabatic \ac{PES} for the NO/Au model with modified coupling.
    }
    \label{fig:extreme-models}
\end{figure}

\begin{figure}
\includegraphics{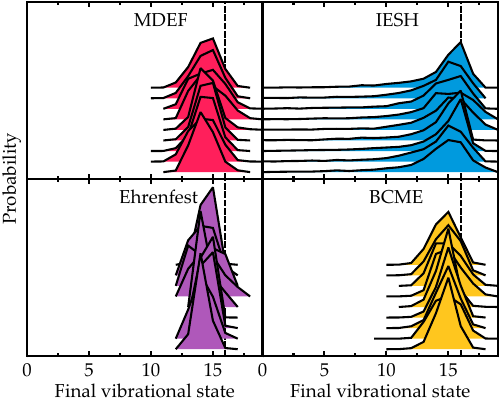}
\caption{
    Final vibrational state distributions for the NO/Au model with increased coupling
    $\Gamma = \SI{15.0}{\eV}$ presented as in \cref{fig:vibrational_distribution_NOAg_16}.
}
\label{fig:vibrational_boxplot_high_gamma}
\end{figure}

\begin{figure}
\includegraphics{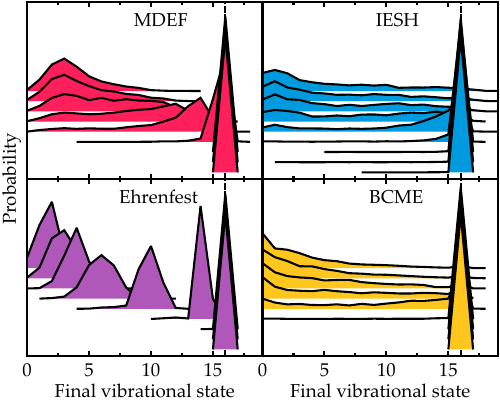}
\caption{
    Final vibrational state distributions for the NO/Au model with reduced coupling
    $\Gamma = \SI{0.15}{\eV}$ presented as in \cref{fig:vibrational_distribution_NOAg_16}.
}
\label{fig:vibrational_boxplot_low_gamma}
\end{figure}

The two models introduced in \cref{sec:models} were chosen to have physically meaningful parameters
to increase the likelihood of correspondence between the model results and physical phenomena.
However, in this section the magnitude of the coupling $\Gamma$ given in \cref{tab:parameters}
is modified to investigate different coupling regimes. 
A direct scaling of $\Gamma$ has the effect of altering the adiabatic ground-state \ac{PES} as we leave $U_0$ and $U_1$ unchanged (\cref{fig:extreme-models}). Therefore, the results of these modified models are not expected to compare to any known realistic system and they  deviate significantly 
from those in \cref{sec:standard-model-results}. Instead, we focus on the effect of $\Gamma$  on the relative agreement between the simulation methods.
To modify the coupling, $\Gamma$ is scaled by a factor of 10 in both directions,
with the high gamma $\Gamma = \SI{15.0}{\eV}$ results shown in \cref{fig:vibrational_boxplot_high_gamma}.
and the low gamma $\Gamma = \SI{0.15}{\eV}$ results shown in \cref{fig:vibrational_boxplot_low_gamma}.
Results are shown only for the modified NO/Au model. As a reminder, a low value of $\Gamma$ means that the molecular state only weakly hybridizes with the continuum of metal states and the impurity state remains a narrow feature in the \ac{DOS}. For very large values of $\Gamma$, the molecular state is broadly hybridised across the electronic density of states and all metal electronic states contain an small admixture of the molecular state $U_1$.

When the coupling is large (\cref{fig:vibrational_boxplot_high_gamma}) all four methods give very similar results,
with only a small amount of vibrational deexcitation for all incidence energies.
For this extreme coupling value the model has entered the adiabatic regime so that each method is expected to perform well.
In fact, the same result is also recovered by adiabatic \ac{MD}.
In the strong coupling regime, all methods are similarly capable to describe the dynamics as the role of nonadiabatic transitions is diminished. 

For the model with small coupling (\cref{fig:vibrational_boxplot_low_gamma}) there is a sudden change in behaviour,
where scattering is vibrationally elastic for low kinetic energies and only becomes inelastic for $E_\text{kin} > \SI{0.4}{\eV}$.
For this model, the usual trend where \ac{MDEF} most closely matches  \ac{BCME} has changed. 
Now \ac{IESH} most closely matches \ac{BCME}.
Both \ac{MDEF} and Ehrenfest are expected to work best in the (quasi-)adiabatic regime, when $\Gamma$ is large,
so it is not surprising that for this reduced $\Gamma$ value they do not fully capture the nonadiabatic energy loss behaviour.

Having observed the results of simulations with artificially modified coupling strength,
the coupling regime of the original models becomes clearer.
For the strong coupling regime, we find good agreement between all methods. For the narrow coupling, we find that MDEF and Ehrenfest perform less well. This suggests that the coupling of the models fitted to the DFT results is in an intermediate regime, where the molecule-metal coupling strength gives similar timescales for nuclear and electronic motion.

To quantify the degree of nonadiabaticity in the model,
we can consider the relative timescales for nuclear and electronic motion.
\cite{douNonadiabaticMolecularDynamics2020}
Although $\Gamma$ depends on the molecular coordinates,
$\hbar/\Gamma$ can be used as a rough metric for the timescale of electronic motion.
For the nuclear motion, the standard harmonic approximation can be extended for
the current model by including the translational kinetic energy $E_\text{kin}$ and
the vibrational state $\nu_i$ to give $\hbar/E_k + 1/\omega\nu_i$ for the nuclear timescale.
By comparing these quantities, we can identify the adiabatic regime,
with fast electronic motion compared to the nuclear motion,
\begin{equation}
\Gamma > \left(\frac{1}{E_\text{kin}} + \frac{1}{\hbar\omega\nu_i}\right)^{-1},
\label{eq:adiabatic-regime}
\end{equation}
and the nonadiabatic regime, where nuclear motion is fast compared to electronic dynamics:
\begin{equation}
\Gamma < \left(\frac{1}{E_\text{kin}} + \frac{1}{\hbar\omega\nu_i}\right)^{-1}.
\label{eq:nonadiabatic-regime}
\end{equation}

\begin{figure}
\includegraphics{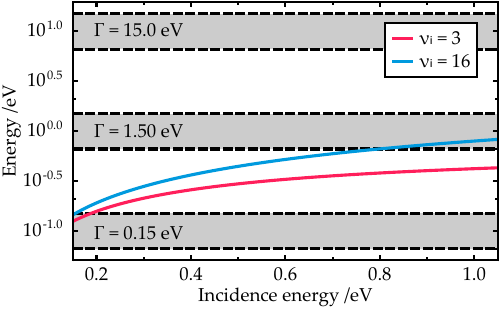}
\caption{
Relative magnitudes of quantities in \cref{eq:adiabatic-regime,eq:nonadiabatic-regime}
for the range of incidence kinetic energies and vibrational states simulated in
\cref{sec:standard-model-results,sec:extreme-coupling}.
The two curves show the right hand side of \cref{eq:adiabatic-regime,eq:nonadiabatic-regime}
as a function of incidence kinetic energy.
The shaded grey regions show the value of $\Gamma$ experienced by the molecule
below heights of \SI{3.5}{\angstrom} above the surface for the three coupling regimes.
}
\label{fig:regimes}
\end{figure}

\Cref{fig:regimes} shows the relative magnitudes of the quantities in
\cref{eq:adiabatic-regime,eq:nonadiabatic-regime}
for the parameters used in \cref{sec:standard-model-results,sec:extreme-coupling}.
As incidence kinetic energy and initial vibrational state increase,
the degree of nonadiabaticity also increases, due to increased speed of the nuclei.
By comparing the relative magnitudes of the nuclear motion shown by the curves,
and the range of explored $\Gamma$ shown by the shaded regions,
it is clear that the high $\Gamma$ model exists in the adiabatic regime
and the low $\Gamma$ model exists in the nonadiabatic regime.
For the physical models ($\Gamma = \SI{1.5}{\eV}$),
at low incidence energy the model appears to exist in the adiabatic regime,
but as the translational energy and vibrational state increase,
the relative timescales of nuclear and electronic motion become comparable.
This suggests a crossover into an intermediate regime,
an observation consistent with the results of the numerical simulations.
Furthermore, for the NO on Au(111) system,
the analysis can be used to justify previous work where it was shown that
\ac{MDEF} can effectively describe low energy scattering but
begins to break down for high vibrational states and increased incidence kinetic energy.
\cite{boxDeterminingEffectHot2021}
The success of the simple metric introduced here implies that for
non-equilibrium scattering problems based on the \ac{NAH}, 
it is possible to inform the choice of simulation method using a small selection of model parameters.

\section{Conclusions}\label{sec:conclusion}

We have introduced two analytic models to study the vibrationally inelastic scattering of an NO molecule on two different metal surfaces, namely Au and Ag.
Using these models, we have assessed the performance of a selection of \ac{MQC} methods,
attempting to bridge the gap between simple harmonic models and full-dimensional simulations that model experiments.
Within the limitations of the models,
we have found that the methods predict similar trends in initial kinetic and vibrational energy dependence,
yet observe consistent variations in the widths of the final vibrational state distributions. Crucially, all models are able to capture important physical trends in initial kinetic and vibrational energy dependence that are consistent with experiment and literature. 

Using \ac{BCME} as a reference method,
we find that \ac{MDEF} is reliably capable of closely matching the result for the physically motivated models (although slightly underestimating the width of the final state distributions).  \ac{IESH} simulations provide relatively good agreement with \ac{BCME} for high vibrational initial states but tend to provide overly broad vibrational energy loss distributions.
In addition, for \ac{IESH} we have introduced a modification of the \ac{EDC} method that is able to improve the results,
suggesting that decoherence effects should be considered when studying molecule-metal scattering. 
By modifying the magnitude of the molecule-metal coupling we are able to establish that the model parameters extracted from previously published \ac{DFT} data 
exist in an intermediate regime such that the timescales for nuclear and electronic motion are comparable.
We have introduced a simple metric that uses the relative magnitudes of the
incidence kinetic energy, initial vibrational state, and molecule-metal coupling
 to identify the regime of nonadiabaticity for models that take the form of the \ac{NAH}.
This metric can be used to inform decisions regarding which \ac{MQC} methods to use in the future.

To build upon this work, the results could be verified by an exact quantum reference such as the \ac{HQME}s
to ensure that \ac{BCME} is indeed a valid reference.
On the topic of decoherence corrections in \ac{IESH} it will be worthwhile to investigate the relative performance of different  decoherence corrections
for a collection of benchmark problems.
To go beyond the simple models investigated here, in the future, it may be possible to parametrize high-dimensional models more closely to ab initio data\cite{ghanInterpretingUltrafastElectron2023}
and make dynamics simulations feasible using machine learning techniques.\cite{zhangEquivariantAnalyticalMapping2022}

It is hoped that this work can be used as a foundation for further tests of \ac{MQC} methods for dynamics at surfaces.
To this end, the models introduced may be used as test systems for methods that
emerge in the future to comprehensively compare their performance or as starting points to explore other effects and parameter regimes.
As our ability to simulate nonadiabatic dynamics of molecules on metal surfaces improves,
we can better explain experimentally observed phenomena and work towards greater control of hot electron effects in chemical dynamics and catalysis at surfaces.

\begin{acknowledgement}
This work was financially supported by the Leverhulme Trust (RPG-2019-078), the UKRI Future Leaders Fellowship program (MR/S016023/1), and a UKRI Frontier research grant (EP/X014088/1). High-performance computing resources were provided via the Scientific Computing Research Technology Platform of the University of Warwick.
\end{acknowledgement}

\section*{Author declarations}
\subsection*{Conflict of interest}
The authors have no financial or non-financial conflicts to disclose.

\subsection*{Author contributions}

\textbf{James Gardner:}
Conceptualization (equal);
data curation (lead);
investigation (lead);
methodology (lead);
software (lead);
visualization (lead);
writing -- original draft (lead);
writing -- review and editing (equal).
\textbf{Scott Habershon:}
Conceptualization (supporting);
supervision (supporting);
writing -- review and editing (supporting).
\textbf{Reinhard J. Maurer:}
Conceptualization (equal);
supervision (lead);
writing -- review and editing (equal).

\section*{Data availability statement}
NQCDynamics.jl is open-source and available at: \url{https://github.com/NQCD/NQCDynamics.jl}. 
Scripts for generating the data and plotting the figures
in this manuscript are available at: \url{https://doi.org/10.5281/zenodo.7973913}.

\bibliography{main}

\end{document}